\documentclass[aps,twocolumn,showpacs,floats]{revtex4}
\usepackage{graphicx}
\usepackage{dcolumn}
\usepackage{bm}
\usepackage{color}
\usepackage{amssymb}

\begin{document}

\title{Critical Temperature of Interacting Bose Gases in Periodic Potentials}

\author{T. T. Nguyen$^{1,2}$, A. J. Herrmann$^{3}$, M. Troyer$^{4}$, and S. Pilati$^{1}$}
\affiliation{$^{1}$The Abdus Salam International Centre for Theoretical Physics, 34151 Trieste, Italy}
\affiliation{$^{2}$SISSA - International School for Advanced Studies, 34136 Trieste, Italy}
\affiliation{$^{3}$Department of Physics, University of Fribourg, 1700 Fribourg, Switzerland}
\affiliation{$^{4}$Theoretische Physik, ETH Zurich, 8093 Zurich, Switzerland}

\begin{abstract}
The superfluid transition of a repulsive Bose gas in the presence of a sinusoidal potential which represents a simple-cubic optical lattice is investigate using quantum Monte Carlo simulations. 
At the average filling of one particle per well the critical temperature has a nonmonotonic dependence on the interaction strength, with an initial sharp increase and a rapid suppression at strong interactions in the vicinity of the Mott transition.
In an optical lattice the positive shift of the transition is strongly enhanced compared to the homogenous gas.
By varying the lattice filling we find a crossover from a regime where the optical lattice has the dominant effect to a regime where interactions dominate and the presence of the lattice potential becomes almost irrelevant.
\end{abstract}

\maketitle


The combined effect of interparticle interactions and external potentials plays a fundamental role in determining the quantum coherence properties of several many-body systems, including He in Vycor or on substrates, paired electrons in superconductors and in Josephson junction arrays, neutrons in the crust of neutron stars~\cite{gandolfi}  and ultracold atoms in optical potentials.
However, even the (apparently) simple problem of calculating the superfluid transition temperature $T_c$ of a dilute homogeneous Bose gas has challenged theoreticians for decades~\cite{lee}. Many techniques have been employed obtaining contradicting results,  differing even in the functional dependence of $T_c$ on the interaction parameter (the two-body scattering length $a$) and in the sign of the shift with respect to the ideal gas transition temperature $T_c^0$ (for a review see Ref.~\cite{andersen}).
In the weakly interacting limit the shift of the critical temperature $\Delta T_c = T_c - T_c^0$ due to interactions has a linear dependence $\Delta T_c/ T_c^0 \simeq c n^{1/3}a$~\cite{baymPRL99,holzmann2001}, where $n$ is the density and the coefficient $c=1.29(5)$ was determined using Monte Carlo simulations of a classical-field model defined on a discrete lattice~\cite{Kashurnikov01,ArnoldMoore01}.
Continuous-space Quantum Monte Carlo simulations of Bose gases with short-range repulsive interactions have shown that this linear form is valid only in the regime $n^{1/3}a \lesssim 0.01$, while at stronger interaction $T_c$ reaches a maximum where $\Delta T_c / T_c^0 \simeq 6.5\%$ and then decreases for $n^{1/3}a \gtrsim 0.2$~\cite{pilati2008}. This suppression of $T_c$ occurs in a regime where universality in terms of the scattering length is lost and other details of the interaction potential become relevant~\cite{giorgini99,pilati2006,pilati2008}.\\
In recent years ultracold atomic gases have emerged as the ideal experimental test bed for many-body theories~\cite{zwerger}.
However, the direct measurement of interactions effects on $T_c$ has been hindered by the presence of the harmonic trap. In the presence of confinement the main interactions effect can be predicted by mean-filed theory and is due to the broadening of the density profile~\cite{stringari}, leading to a suppression of $T_c$. Deviations from the mean-field prediction and effects due to critical correlations have been measured in Refs.~\cite{hadzibabic2011,hadzibabic2011bis}. 
A major breakthrough has been achieved recently with the realization of Bose-Einstein condensation in quasi-uniform trapping potentials~\cite{hadzibabic2013}.
This setup allows a more direct investigation of critical points where a correlation length diverges and the arguments based on the local density approximation become invalid.

\begin{figure}
\begin{center}
\includegraphics[width=1.0\columnwidth]{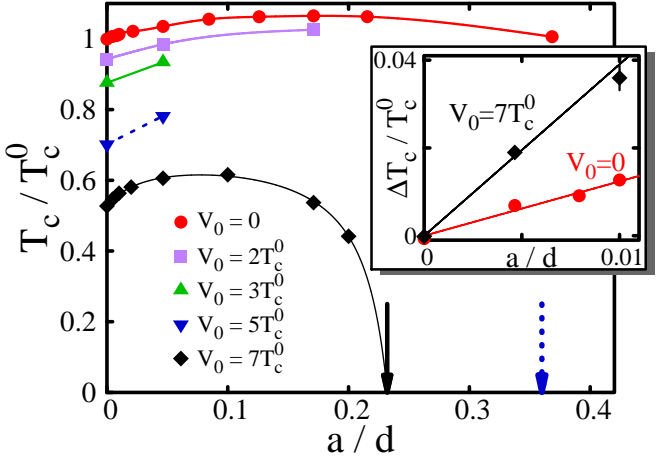}
\caption{Main panel: Superfluid transition temperature $T_c/T_c^0$ as a function of the scattering length $a$, for different intensities of the optical lattice $V_0$. The filling is fixed at $nd^3=1$ ($n$ is the density and $d$ the lattice spacing). The vertical arrows indicate the $T=0$ Mott insulator transition for $V_0=7T_c^0$ (black) and $V_0=5T_c^0$ (dashed blue)~\cite{pilati2012}. The lines are guides to the eye.
Inset: Shift of $T_c$ with respect to the value at $a = 0$. The solid lines are linear fits. 
The transition temperature of the homogeneous noninteracting gas ($V_0=0$ and $a=0$) is $T_c^0 \cong 0.671E_R$, where $E_R$ is the recoil energy.}
\label{tc}
\end{center}
\end{figure}

\begin{figure}
\begin{center}
\includegraphics[width=1.0\columnwidth]{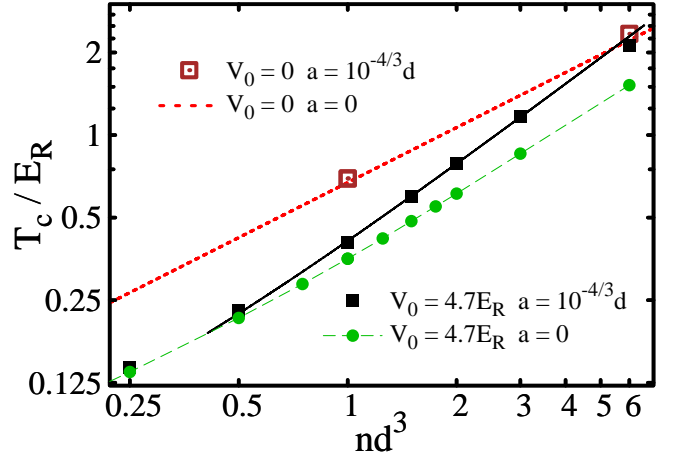}
\caption{ Critical temperature $T_c/E_R$ as a function of the filling factor $nd^3$ for fixed interaction strength $a/d$. The red-dashed line is the critical temperature of the homogeneous noninteracting Bose gas $T_c^0\propto n^{2/3}$. The black solid line is a linear fit on $T_c$ of the interacting gas in the optical lattice in the range $0.5 \leq nd^3 \leq 3$. The long-dashed green line is a guide to the eye.}
\label{tcnd3}
\end{center}
\end{figure}

%
The superfluid transition in the presence of periodic potentials is even more complex than in homogeneous systems due to the intricate interplay between interparticle interactions and the external potential and to the role of commensurability.
In this Letter we employ unbiased quantum Monte Carlo methods to determine the critical temperature of a 3D repulsive Bose gas in the presence of a simple-cubic optical lattice with spacing $d$.
We find that at the integer filling $nd^3 = 1$ (an average density of one bosons per well of the external field) the critical temperature $T_c$ has an intriguing nonmonotonic dependence on the interaction strength (parametrized by the ratio $a / d$) with an initial increase and a rapid suppression at strong interaction in the vicinity of the Mott insulator quantum phase transition. Counterintuitively, the initial increase is stronger in the optical lattice than in homogenous systems (see Fig.~\ref{tc}).
By varying the filling $nd^3$ at fixed interaction parameter $a / d$, we observe a crossover from a low-density regime where the effect of interactions is marginal and $T_c$ is essentially the same as in the noninteracting case, to a regime at large $nd^3$ where the role of interactions is dominant while the effect of the optical lattice becomes almost negligible and $T_c$ approaches the homogeneous gas value. In the crossover region we observe that $T_c$ varies linearly with $nd^3$ (see Fig.~\ref{tcnd3}).\\
In our simulations we consider a gas of spinless bosons described by the Hamiltonian:
\begin{equation}
H=\sum_{i=1}^N \left(-\frac{\hbar^2}{2m}\nabla_i^2+V({\bf r}_i)\right)+\sum_{i<j}v(|{\bf r}_i-{\bf r}_j|) \;,
\label{hamiltonian}
\end{equation}
where $\hbar$ is the reduced Planck constant, $m$ the particle mass and the vectors ${\bf r}_i$ denote the coordinates of the $N$ particles labeled by the index $i$. The pairwise interparticle interactions is modeled by the hard-sphere potential: $v(r)=+\infty$ if $r<a$ and zero otherwise, where the hard-sphere diameter $a$ corresponds to the $s$-wave scattering length. $V(\mathbf{r})=V_0\sum_{\alpha=x,y,z}\sin^2\left(\alpha\pi/d\right)$ is a simple-cubic optical lattice potential with spacing $d$ and intensity $V_0$, which we shall express in units of $T_c^0\cong 3.3125\hbar^2n^{2/3}/m$ or recoil energy $E_R=  \hbar^2\pi^2/(2md^2)$ (we set the Boltzmann constant $k_B = 1$). The bosons are in a cubic box of volume $V=(N_s d)^3$ (where $N_s$ is an integer) with periodic boundary conditions.\\
%
To simulate the thermodynamic properties of the Hamiltonian~(\ref{hamiltonian}) we employ the Path Integral Monte Carlo (PIMC) method~\cite{Ceperley95}. This technique provides unbiased estimates of thermal averages of physical quantities using the many-particle configurations ${\bf R}=({\bf r}_1,...,{\bf r}_N)$ sampled from a probability distribution proportional to the density matrix $\rho({\bf R},{\bf R},T)=\langle{\bf R}|e^{-H/T}|{\bf R}\rangle$ at the temperature $T$. 
We are interested in the superfluid fraction $\rho_S/\rho$ (where $\rho = mn$ is the total mass density), obtained from the winding number estimator~\cite{PollockCeperley87}, and in the one-body density matrix $n_1({\bf r},{\bf r}')=\left< \psi^\dag({\bf r}) \psi({\bf r}')\right>$, where $\psi^\dag({\bf r})$ ($\psi({\bf r})$) is the bosonic creation (annihilation) operator. These quantities are efficiently evaluated in PIMC simulations if configuration sampling is performed using the worm algorithm~\cite{boninsegni06PRE}. 
%
For more details on the implementation of the PIMC algorithm, see Refs.~\cite{pilati2010,boninsegni06PRE,pilati2006} and the Supplemental Material~\cite{supplemental}.\\
%
%
%
%
\begin{figure}
\begin{center}
\includegraphics[width=1.0\columnwidth]{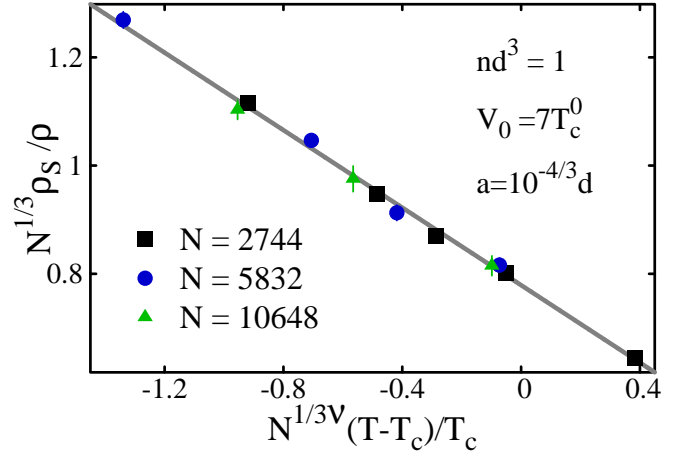}
\caption{Scaled superfluid fraction as a function of the scaled reduced temperature. Data obtained for different particles numbers $N$ collapse on top of the universal scaling function $f(x)$, see eq.~(\ref{scaling}) (thick gray line).}
\label{figscaling}
\end{center}
\end{figure}
%
The critical temperature $T_c$ is determined from a finite-size scaling analysis of $\rho_s/\rho$ using the scaling Ansatz~\cite{PollockRunge92}:
\begin{equation}
N^{1/3}\rho_S(t,N)/\rho=f(tN^{1/3\nu})= f(0)+f^\prime(0)tN^{1/3\nu}+... \;.
\label{scaling}
\end{equation}
Here, $t=(T-T_c)/T_c$ is the reduced temperature, $\nu$ is the critical exponent of the correlation length $\xi \sim t^{-\nu}$, and $f(x)$ is a universal analytic function which allows for a linear expansion close to $x = 0$.
We obtain $T_c$, $f(0)$, $f^\prime(0)$ and $\nu$ from a best fit analysis of PIMC data obtained with different system sizes~\cite{notesize}.
In agreement with the scaling Ansatz~(\ref{scaling}), the PIMC results for the rescaled superfluid fraction $N^{1/3} \rho_S/\rho$ plotted as a function of the rescaled reduced temperature $N^{1/3\nu}t$ collapse on top of a universal scaling function $f(x)$ (see Fig.~\ref{figscaling}).
The values of $\nu$ obtained from the best-fit analysis are consistent with the critical exponent of the 3DXY model $\nu \simeq 0.67$~\cite{campostrini} in the interacting case, and with $\nu = 1$  (corresponding to the gaussian complex field model) in the noninteracting case~\cite{noteuniversalityclass}.
%
%
For selected values of $V_0$, $a/d$ and $nd^3$~\cite{selectedvalues} we determine $T_c$ also by calculating the fraction of particles with zero momentum $n_0/n$ (sometimes referred to as  coherent fraction), which can be extracted from the long-distance behavior of the one-body density matrix $n_1({\bf r},{\bf r}')$ (see~\cite{supplemental}).
%
%
%
In the noninteracting case we obtain $T_c$ also by calculating via exact diagonalization the condensate fraction $n_C/n$, {\it i.~e.} the fraction of particles in the Bloch state with zero quasi-momentum~\cite{supplemental,bloch}.
All methods we employ to determine $T_c$ provide predictions which are consistent within statistical errors.
%
%
%
\begin{figure}
\begin{center}
\includegraphics[width=1.0\columnwidth]{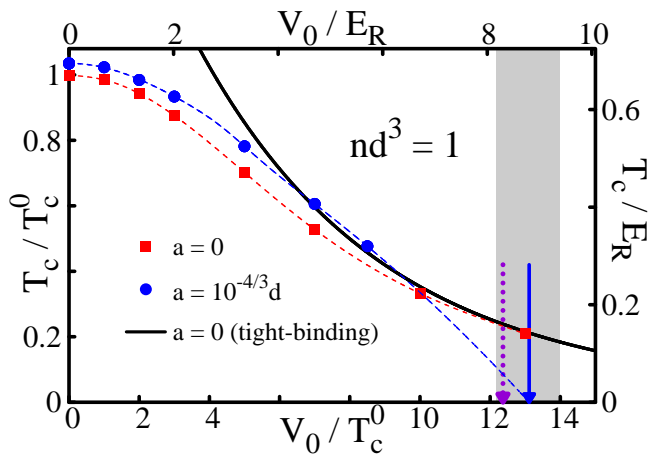}
\caption{Critical temperature as a function of the optical lattice intensity. 
The Mott insulator transition at $T=0$ is indicated by the solid blue vertical arrow (from Ref.~\cite{pilati2012}), by the the dashed violet arrow (obtained via approximate mapping to the Bose-Hubbard model~\cite{capogrosso,zoller}) and by the shaded gray area (experimental result of Ref.~\cite{nagerl} with errorbar~\cite{noteexp}). 
Thin dashed lines are guides to the eye.}
\label{tcVo}
\end{center}
\end{figure}
\begin{figure}
\begin{center}
\includegraphics[width=1.0\columnwidth]{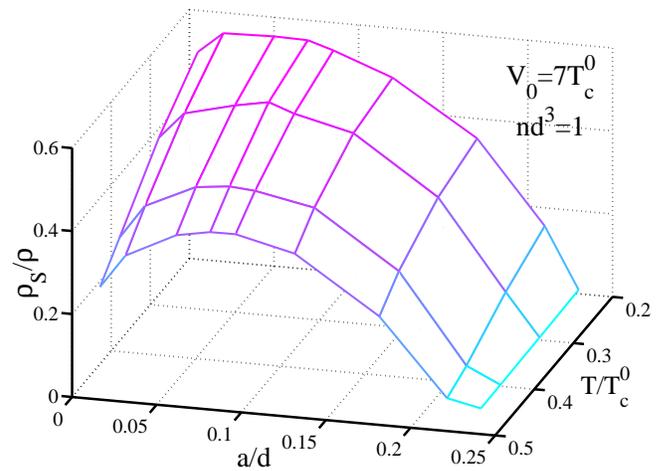}
\caption{Superfluid fraction as a function of the temperature $T/T_c^0$ and the interaction strength $a/d$.}
\label{rhoS}
\end{center}
\end{figure}
In Fig.~\ref{tcVo} we show the dependence of $T_c$ on the strength of the optical lattice potential $V_0$ at integer filling $nd^3=1$. Both in the interacting and in the noninteracting case $T_c$ monotonically decreases as $V_0$ increases. In moderately intense lattices as the one considered in this work thermal excitations populate higher Bloch bands making the single-band approximation invalid.
Indeed we observe that the noninteracting critical temperature converges to the tight-binding result~\cite{capogrosso} only for $V_0 \gtrsim 12T_c^0$. Then it vanishes asymptotically in the large $V_0$ limit.  In the interacting case $T_c$ is increased compared to the noninteracting case in shallow optical lattices. As the lattice gets deeper $T_c$ rapidly decreases approaching zero at the quantum phase transition to the Mott insulator. In the proximity of the quantum critical point, finite-temperature PIMC simulations become impractical due to critical slowing-down; even so the trend of our data at intermediate $T$ is consistent with the critical point predicted by previous Monte Carlo simulations of the ground state of the Hamiltonian~(\ref{hamiltonian})~\cite{pilati2012} and with the experimental result of Ref.~\cite{nagerl}.\\
The nonmonotonic dependence of $T_c$ as a function of the interaction parameter is highlighted in Fig.~\ref{tc}. Interactions effects are larger in an optical lattice than in the homogenous gas ($V_0 = 0$). If we assume a linear dependence $\Delta T_c/T_c^0 = c n^{1/3} a = c a/d$ (here we consider the shift $\Delta T_c$ from the critical point at the given $V_0$ and $a = 0$), a best fit analysis in the range $0 \leq a /d\leq 0.01$ provides the coefficients $c = 3.9(3)$ for $V_0=7T_c^0$ and $c = 1.24(7)$ for $V_0 = 0$ (see inset in Fig.~\ref{tc}). These results indicate a cooperative interplay between interactions and external potential.
The superfluid density also shows a nonmonotonic dependence on $a/d$, even well below the critical temperature (see Fig.~\ref{rhoS}).\\
Fig.~\ref{tcnd3} displays how $T_c$ varies with the lattice filling if the interaction strength is fixed at $a / d=10^{-4/3}$ and the optical lattice intensity at $V_0=7T_c^0$ . At low filling ($nd^3 \approx 0.25$) the critical temperature is almost unaffected by interactions. On the other hand, at high filling ($nd^3 \approx 6$) the role of interactions is dominant while the optical lattice becomes unimportant and $T_c$ approaches the transition of the homogeneous system. In the crossover region $0.5\leq nd^3 \leq 3$ the dependence of $T_c$ on the density is accurately described by a simple linear fitting function $T_c(nd^3) = E_R\left[0.376(2)nd^3 + 0.036(4)\right]$.
It is worth noticing that in the optical lattice interactions can induce important changes of $T_c$, up to $40\%$ at $nd^3 = 6$, much larger than in the homogeneous case.
We explain this intriguing behavior of $T_c$ as a consequence of the screening of the external potential due to the interactions. This screening inhibits the suppression of $T_c$ which would otherwise be induced by the optical lattice if the particles were noninteracting.
%
%

Both the sharp positive shift of $T_c$ at fixed filling $nd^3 = 1$ and the linear dependence on $nd^3$ at fixed interaction strength take place in a regime of small values of the diluteness parameter $na^3 \lesssim 5 \cdot 10^{-4}$. In this region universality in terms of the scattering length is preserved, both in absence of external potentials~\cite{giorgini99} and in the optical lattice~\cite{pilati2012}. Details of our model interparticle potential other than $a$ ({\it e.g.}, the effective range and the scattering lengths in higher partial waves) are irrelevant,  hence our results quantitatively describe experiments performed with ultracold atomic gases in which the interaction strength is tuned using broad Feshbach resonances.
%
%

In conclusion, we have investigated the combined effect of interactions and external periodic potentials on the superfluid transition in a 3D Bose gas.
Previous approximate theoretical studies addressed the onset of superfluidity in weak unidirectional optical lattices~\cite{zobay}, and the mean-field suppression of $T_c$ in combined harmonic plus optical-lattice potentials~\cite{blakie}.
The determination of $T_c$ in extended systems is a highly nonperturbative problem that can be rigorously solved only using unbiased quantum many-body techniques such as the PIMC method employed in this work.
PIMC simulations have already been applied to investigate the superfluid transition in liquid $^4$He~\cite{Boninsegni06}, in dilute homogenous Bose gases~\cite{Gruter97,NhoLandau04,pilati2008}, in dipolar systems~\cite{filinov} and in disordered Bose gases~\cite{carleo,pilati2009}.\\
So far, the theoretical studies and the experiments performed on optical lattice systems have been focused on the suppression of $T_c$~\cite{trotzky} and on the localization transition~\cite{greiner2002quantum,nagerl} which take place in deep lattices and strong interatomic interaction. 
In this work we show that correlations have a more intriguing effect on the quantum-coherence properties than what was previously assumed.
In the regime of weak interactions the superfluid fraction and the critical temperature are enhanced by interparticle repulsion.
Counterintuitively, in commensurate lattices of moderate intensity the upward shift of $T_c$ is even more pronounced compared to the weak effect observed in homogeneous systems. This shift of $T_c$ further increases when the filling factor is tuned above unity. In this regime the presence of the periodic potential becomes essentially irrelevant.
The recent realization of quasi-uniform trapping potentials~\cite{hadzibabic2013} for atomic clouds gives strong hope that these findings can be observed in experiments.

We acknowledge support by the Swiss National Science Foundation.



\newpage
\begin{center}
\bf{Supplemental Material for\\
Critical Temperature of Interacting Bose Gases in Periodic Potentials}

\author{T. T. Nguyen$^{1,2}$, A. J. Herrmann$^{3}$, M. Troyer$^{4}$, and S. Pilati$^{1}$}
\affiliation{$^{1}$The Abdus Salam International Centre for Theoretical Physics, 34151 Trieste, Italy}
\affiliation{$^{2}$SISSA - International School for Advanced Studies, 34136 Trieste, Italy}
\affiliation{$^{3}$Department of Physics, University of Fribourg, 1700 Fribourg, Switzerland}
\affiliation{$^{4}$Theoretische Physik, ETH Zurich, 8093 Zurich, Switzerland}
\end{center}


To determine the superfluid fraction $\rho_S/\rho$ and the coherent fraction $n_0/n$ of interacting Bose gases we employ the Path Integral Monte Carlo (PIMC) method~\cite{SCeperley95}.
In PIMC simulations the density matrix at temperature $T$ is obtained via Trotter discretization from an appropriate approximate form valid at the higher temperature $T\cdot M$, where the integer $M$ is the Trotter number. For the one-body term of the Hamiltonian defined in eq.~(1) of the main text we use the symmetrized primitive approximation, while for the pair-wise additive potential we use the pair-product approximation~\cite{SCeperley95,Skrauth}. We approximate the pair density matrix of the hard-sphere potential using the Cao-Berne analytical formula~\cite{Scaoberne}. This approximation was found to be comparably accurate as the exact pair density matrix obtained numerically via partial wave expansion~\cite{Spilati2006}.  The PIMC method is exact in the limit $M\rightarrow \infty$. To analyze the possible bias due to a finite Trotter number we performed benchmark simulation with up to $M=164$. At the intermediate temperatures considered in this work, values of the Trotter number equal to $M= 32$ or to $M=64$ are found to provide estimates of $\rho_S/\rho$ and $n_0/n$ which coincide with the extrapolation to $M\rightarrow \infty$ within our statistical uncertainty.
For more details on the computational method, see Refs.~\cite{Spilati2006,Spilati2010,Sboninsegni06PRE}.\\
%
We calculate $\rho_S/\rho$ from the winding number estimator~\cite{SPollockCeperley87}, while $n_0/n$ is obtained from the asymptotic value of the bulk-averaged one-body density matrix: $n_0/n = \lim_{\left|{\bf s}\right| \rightarrow \infty} N^{-1} \int \mathrm{d}{\bf r} n_1\left({\bf r}+{\bf s},{\bf r}\right)$, where the integrand is averaged over the solid angle of ${\bf s}$.
The coherent fraction is the squared modulus of the order parameter that characterizes the superfluid transition~\cite{Skhuang}. For $V_0 = 0$ it coincides with the condensate fraction~\cite{Smuller,Sastra}.
The rescaled coherent fraction $N^{(1+\eta)/3}n_0/n$, involving the critical exponent of the correlation function $\eta$, follows a universal scaling law analogous to eq.~(2), allowing us to determine $T_c$ from a finite-size scaling analysis as in the case of $\rho_S/\rho$. We employ the predictions $\eta \simeq 0.038$~\cite{Scampostrini} for $a > 0$ and $\eta = 0$~\cite{Skhuang} for $a = 0$, corresponding to the universality classes of the 3D XY and the gaussian complex-field models, respectively.\\
%
In the noninteracting case $a=0$ we determine the critical temperature also by calculating the condensate fraction $n_C/n$, {\it i.~e.} the fraction of particles in the lowest-energy single-particle eigenstate.
We obtain the single-particle spectrum by solving the following single-particle Schr\"odinger equation in a 1D box of size $L=N_Sd$ with periodic boundary conditions~\cite{Sbloch}:
\begin{equation}
\label{schroedinger}
\left[\frac{-\hbar^2}{2m} \frac{\partial^2}{\partial x^2} + V_0 \sin^2\left(x \pi/d\right)\right]\phi_{q_x}^{(n_x)}(x) = E_{q_x}^{(n_x)} \phi_{q_x}^{(n_x)}(x);
\end{equation}
the eigenstates are the Bloch functions $\phi_{q_x}^{(n_x)}(x) = \exp\left(iq_x x/\hbar\right)u_{q_x}^{(n_x)}(x)$, where $n_x=1,2,\dots$ is the Band index~\cite{Sashcroft}. The quasi-momentum can take the values $q_x =i2\pi/L$, with the integer $i$ in the range $i = -N_S/2 < i \leq N_S/2$.
The simple-cubic optical lattice is separable, thus the 3D eigenvalues can be written as $E_{\bf q}^{({\bf n})}=E_{q_x}^{(n_x)}+E_{q_y}^{(n_y)}+E_{q_z}^{(n_z)}$, with the quasi-momentum ${\bf q} = (q_x,q_y,q_z)$ and the band index ${\bf n} = (n_x,n_y,n_z)$. The chemical potential $\mu$ is fixed by the normalization condition $N = \sum_{\bf n}\sum_{\bf q} N _{\bf q}^{({\bf n})}$, where the mean eigenstate occupations are given by the Bose distribution $N _{\bf q}^{({\bf n})} = 1/\left(\exp(E_{\bf q}^{({\bf n})} -\mu)/T -1\right)$.
We determine the Bose-Einstein critical temperature below which the condensate fraction $n_C/ n=N_{0,0,0}^{(0,0,0)}/N$ remains finite in the thermodynamic limit~\cite{Spethick}. 
In the $V_0=0$ case the result coincides with $T_c^0\cong3.3125\hbar^2n^{2/3}/m$. We recall that an ideal Bose-Einstein condensate is an equilibrium superfluid, even though it does not satisfy the Landau criterion~\cite{Sblatt}.\\
The three methods we employ to determine $T_c$, namely the two based on the PIMC estimates of $\rho_S/\rho$ and $n_0/n$ and the one based on the exact calculation of $n_C/n$ (in the $a=0$ case), provide predictions which coincide within our statistical uncertainty.


\begin{thebibliography}{99}
\bibitem{noteuniversalityclass} However, we notice that $T_c$ is not very sensible to the exact value of $\nu$, and we obtain compatible estimates of $T_c$ by using $\nu$ as a fitting parameter or by fixing its value to the universal predictions given in the text.
\bibitem{noteexp} We obtained the value at $a/d = 10^{-4/3}$ by interpolating $6$ experimental data in the range $0.0053 \leq a/d \leq 0.053$.
The width of the gray area is the errorbar of the experimental point at $a/d = 0.043$.
\end{thebibliography}
\end{document}